\documentclass[showpacs,epsf,twocolumn]{revtex4-1}
\usepackage{float}
\usepackage{color,graphicx}
\graphicspath{ {images/} }
\usepackage{hyperref}
\begin{document}

\title{Probing the Fermi surface and magnetotransport properties in MoAs$_{2}$}

\author{Ratnadwip Singha}
\affiliation{Saha Institute of Nuclear Physics, HBNI, 1/AF Bidhannagar, Calcutta 700 064, India}
\author{Arnab Pariari}
\affiliation{Saha Institute of Nuclear Physics, HBNI, 1/AF Bidhannagar, Calcutta 700 064, India}
\author{Gaurav Kumar Gupta}
\affiliation{Department of Physics, Indian Institute of Science, Bangalore 560 012, India}
\author{Tanmoy Das}
\affiliation{Department of Physics, Indian Institute of Science, Bangalore 560 012, India}
\author{Prabhat Mandal}
\affiliation{Saha Institute of Nuclear Physics, HBNI, 1/AF Bidhannagar, Calcutta 700 064, India}
\email{prabhat.mandal@saha.ac.in}
\date{\today}

\begin{abstract}
Transition metal dipnictides (TMDs) have recently been identified as possible candidates to host topology protected electronic band structure. These materials belong to an isostructural family and show several exotic transport properties. Especially, the large values of magnetoresistance (MR) and carrier mobility have drawn significant attention from the perspective of technological applications. In this report, we have investigated the magnetotransport and Fermi surface properties of single crystalline MoAs$_{2}$, another member of this group of compounds. Field induced resistivity plateau and a large MR have been observed, which are comparable to several topological systems. Interestingly, in contrast to other isostructural materials, the carrier density in MoAs$_{2}$ is quite high and shows single-band dominated transport. The Fermi pockets, which have been identified from the quantum oscillation, are largest among the members of this group and have significant anisotropy with crystallographic direction. Our first-principles calculations reveal a substantial difference between the band structures of MoAs$_{2}$ and other TMDs. The calculated Fermi surface consists of one electron pocket and another 'open-orbit' hole pocket, which has not been observed in TMDs so far.
\end{abstract}

\maketitle

\section{Introduction}

The emergence of topological materials in condensed matter physics, has introduced a completely new perspective, where the materials are characterized in terms of the topology of their electronic band structure and the physics is dictated by the inherent symmetries of the system. These materials are the subject of extensive theoretical and experimental studies and continue to reveal new exotic topological phases of matter. While the discovery of topological insulators (TIs) marks the beginning \cite{Zhang,Xia,Chen}, subsequent realizations of Dirac, Weyl and nodal line semimetals \cite{Liu1,Liu2,Lv,Xu,Bian,Schoop} have further enriched this field of research. Conducting surface and insulating bulk states are the distinctive features of TIs, whereas bulk conduction band and valence band cross at either fourfold/twofold-degenerate discrete points (Dirac/Weyl points) or along one-dimensional line (nodal line) in topological semimetals (TSMs). In the vicinity of these band crossings, the dynamics of the quasi-particle excitations are described by relativistic equations of motion. Apart from the opportunity to explore the novel physics of relativistic particles, topological systems have also caught attention due to their unique transport properties \cite{Ali,Shekhar,Singha}, which can have a huge impact in technological applications \cite{Wolf,Lenz}. However, the real challenge lies in finding new materials, which are suitable for both basic research and technological use.

The family of transition metal dipnictides [$XPn_{2}$ ($X$ = Ta, Nb; $Pn$ = P, As, Sb)], has recently been proposed as ideal candidates for investigating the topology protected electronic systems \cite{Xu2}. These materials possess identical electronic band structure and host multiple band anticrossings near the Fermi level. In absence of spin-orbit coupling (SOC), these anticrossings form a nodal-line in the \textbf{k}-space. However, with the inclusion of SOC, the nodal line is gapped out, leading to only isolated electron and hole pockets. Several magnetotransport studies have been reported for different members of this family \cite{Wang,Shen,Yuan,Wang2,Li}. Though the results have been seen to vary slightly for different systems, they all show low-temperature resistivity saturation, large magnetoresistance (MR) and high carrier mobility, which are some of the characteristics of TSMs. On the other hand, in these reports \cite{Wang,Shen,Yuan,Wang2,Li}, the magnetotransport properties are often attributed to the compensated electron-hole density rather than any non-trivial band topology.

Herein, we have investigated the transport properties of single crystalline MoAs$_{2}$, another member of the $XPn_{2}$ family. We have obtained a large and anisotropic MR and robust low temperature resistivity plateau, in spite of quite high carrier density and single-band dominated transport. The Fermi surface of the material is analyzed from de Haas-van Alphen (dHvA) oscillation. The band structure calculation reveals a large electron-type Fermi pocket along with another 'open-orbit' hole-type Fermi surface.  We have also identified a suspected Weyl cone well above the Fermi energy.

\section{Single crystal growth and experimental details}

The single crystals of MoAs$_{2}$ were grown in iodine vapor transport method \cite{Murray}. At first, the polycrystalline powder was synthesized from elemental Mo (Alfa Aesar 99.95\%) and As (Alfa Aesar 99.9999\%) in a evacuated quartz tube at 950$^{\circ}$C for 5 days. The powder along with iodine (5 mg/cc.) were sealed in another quartz tube under vacuum and put in a gradient furnace for 7 days. The hotter end of the tube, containing the polycrystalline powder, was kept at 950$^{\circ}$C, whereas the other end was maintained at 900$^{\circ}$C. Needle like single crystals with typical dimensions 1$\times$0.8$\times$0.4 mm$^{3}$ were obtained at the cold end. The crystals were characterized using powder X-ray diffraction technique in a Rigaku X-ray diffractometer (TTRAX III). The transport measurements were performed in a 9 T physical property measurement system (Quantum Design) by using ac-transport option and rotating sample holder. The electrical contacts were made using gold wires and conducting silver paint (Leitsilber 200N) in a four-probe configuration. Magnetization measurements were done in 7 T SQUID-VSM MPMS3 (Quantum Design).

\section{Computational details}

Band structure calculations are performed using density functional theory within the Local Density Approximation (LDA) exchange correlation as implemented in Vienna ab-initio simulation package (VASP) \cite{Kresse1}. Projected augmented-wave (PAW) pseudo-potentials are used to describe the core electron in the calculation \cite{Kresse2}. LDA+U method is used to deal with the strong correlation in this material with the standard value of U=2.4 eV on the correlated Mo-4\textit{d} orbitals. The electronic wavefunction is expanded using plane wave up to a cutoff energy of 224.584 eV. Brillouin zone sampling is done by using a (8$\times$8$\times$8) Monkhorst-Pack \textbf{k}-grid. Both atomic position and cell parameters are allowed to relax, until the forces on each atom are less than 0.01 eV/{\AA}.

\section{Results and discussions}

MoAs$_{2}$ crystallizes in OsGe$_{2}$-type structure with monoclinic space group \textit{C12/m1} \cite{Jensen}. The primitive unit cell contains two inequivalent Mo atoms. The crystal structure is shown in Fig. 1(a). Each molybdenum atom is surrounded by six arsenic atoms, which form a trigonal prism. A typical single crystal of MoAs$_{2}$ is shown in the inset of Fig. 1(b) with appropriate length scale. The needle-like crystals of $XPn_{2}$ family, preferably grow along b-axis \cite{Shen}. Here, the top facet of the crystal corresponds to (0 0 1) plane, whereas inclined side facet is the (2 0 1) plane. The long edge, shared between these two crystal planes, indicate the direction of the crystallographic b-axis \cite{Shen}. For simplicity, we have discussed the results along three mutually perpendicular crystallographic directions, b, c [perpendicular to (0 0 1) plane] and b$\times$c-axes, throughout this report. The X-ray diffraction (XRD) spectrum of powdered crystals is shown in Fig. 1(b). The obtained spectrum confirms the phase purity of the grown crystals and absence of any impurity state. The XRD pattern has been analyzed by Rietveld structural refinement using FULLPROF software package. The refined parameters, a=9.062(5){\AA}, b=3.296(2){\AA}, c=7.715(4){\AA} and $\beta$=119.3(2)$^{\circ}$, are in agreement with the earlier report \cite{Jensen}.

\begin{figure}
\includegraphics[width=0.5\textwidth]{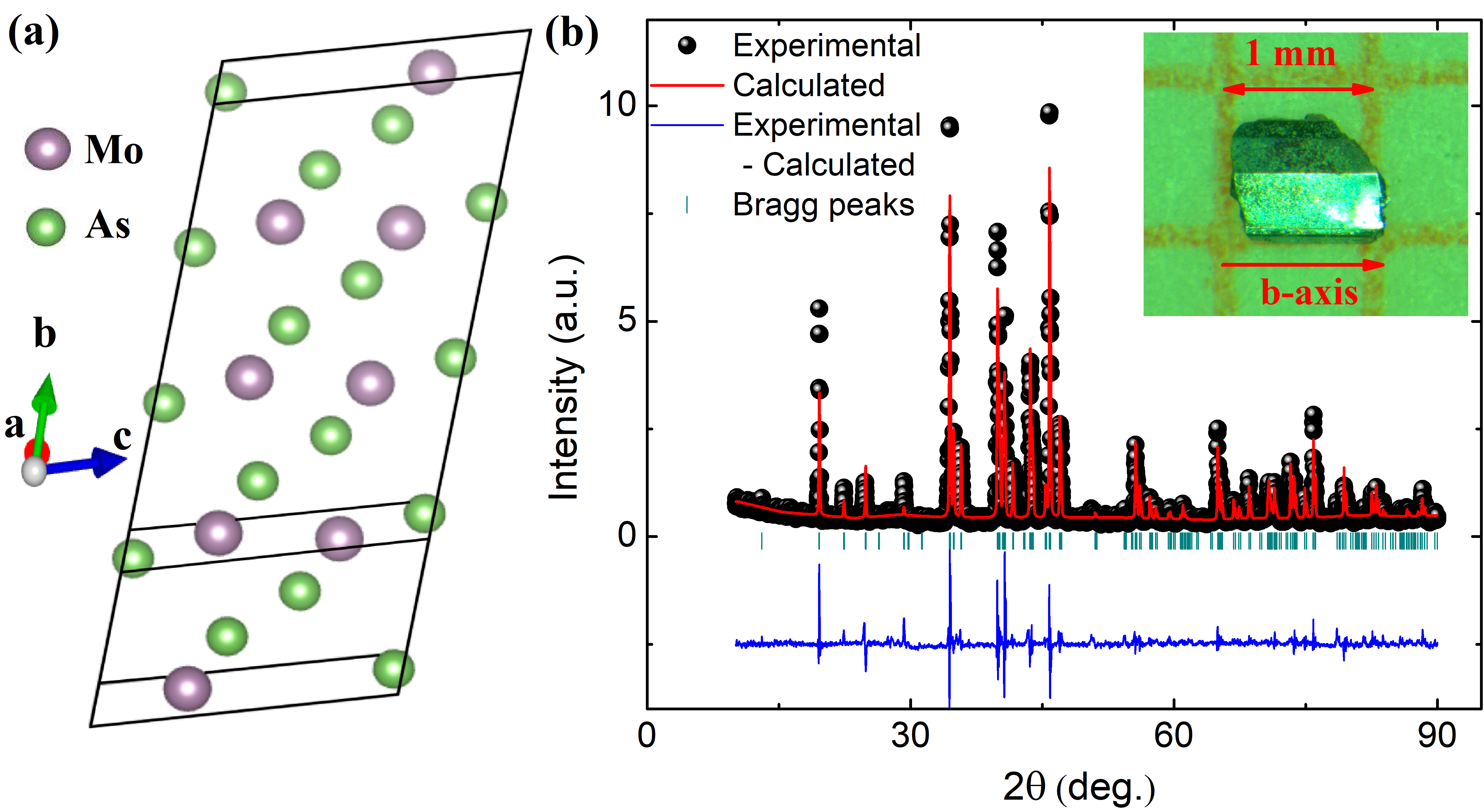}
\caption{(a) Crystal structure of MoAs$_{2}$. (b) X-ray diffraction spectra of the powdered MoAs$_{2}$ single crystals. Inset shows a typical crystal grown along crystallographic b-axis.}
\end{figure}

The resistivity measurements on as grown crystals have been done with current along b-axis. As illustrated in Fig. 2(a), the temperature dependence of resistivity [$\rho_{xx}$(T)] shows metallic character throughout the measured temperature range. $\rho_{xx}$ becomes as small as $\sim$0.35 $\mu\Omega$ cm at 2 K and yields a residual resistivity ratio [RRR=$\rho_{xx}$(300 K)/$\rho_{xx}$(2 K)] $\sim$312. Very low residual resistivity and high RRR are the clear signatures of the high quality of the single crystals. While the resistivity decreases linearly with temperature from 300 K, the experimental data can be fitted well with $\rho_{xx}(T)=a+bT^n$ ($n\sim$3) type relation below 75 K [Fig. 2(a) inset]. When the magnetic field is applied, the low-temperature resistivity shows an enhancement followed by a saturation behavior. This4 field induced metal to semiconductor-like crossover and resistivity plateau have been seen to be generic features of TSMs \cite{Ali,Singha,Shekhar}. Though two different mechanisms based on the field-induced gap opening model \cite{Khveshchenko} and Kohler scaling analysis \cite{Wang3,Singha2} have been proposed, the actual origin of such behavior is yet to be established unambiguously. Nevertheless, two characteristic temperatures ($T_{m}$ and $T_{i}$) can be identified from the $\partial\rho_{xx}/\partial T$ curves in Fig. 2(b). $T_{m}$ is the crossover temperature and indicated by the zero-crossing point of the $\partial\rho_{xx}/\partial T$ curve, whereas $T_{i}$ is the point of minima in $\partial\rho_{xx}/\partial T$ vs. $T$ plot. $T_{i}$ corresponds to the inflection point in $\rho_{xx}(T)$, below which resistivity saturation starts to occur. As shown in the inset of Fig. 2(b), $T_{m}$ increases monotonically with applied field strength and follows a ($B-B_{0}$)$^{1/3}$ type relation. On the other hand, $T_{i}$ shows a weak field dependence.

\begin{figure}
\includegraphics[width=0.33\textwidth]{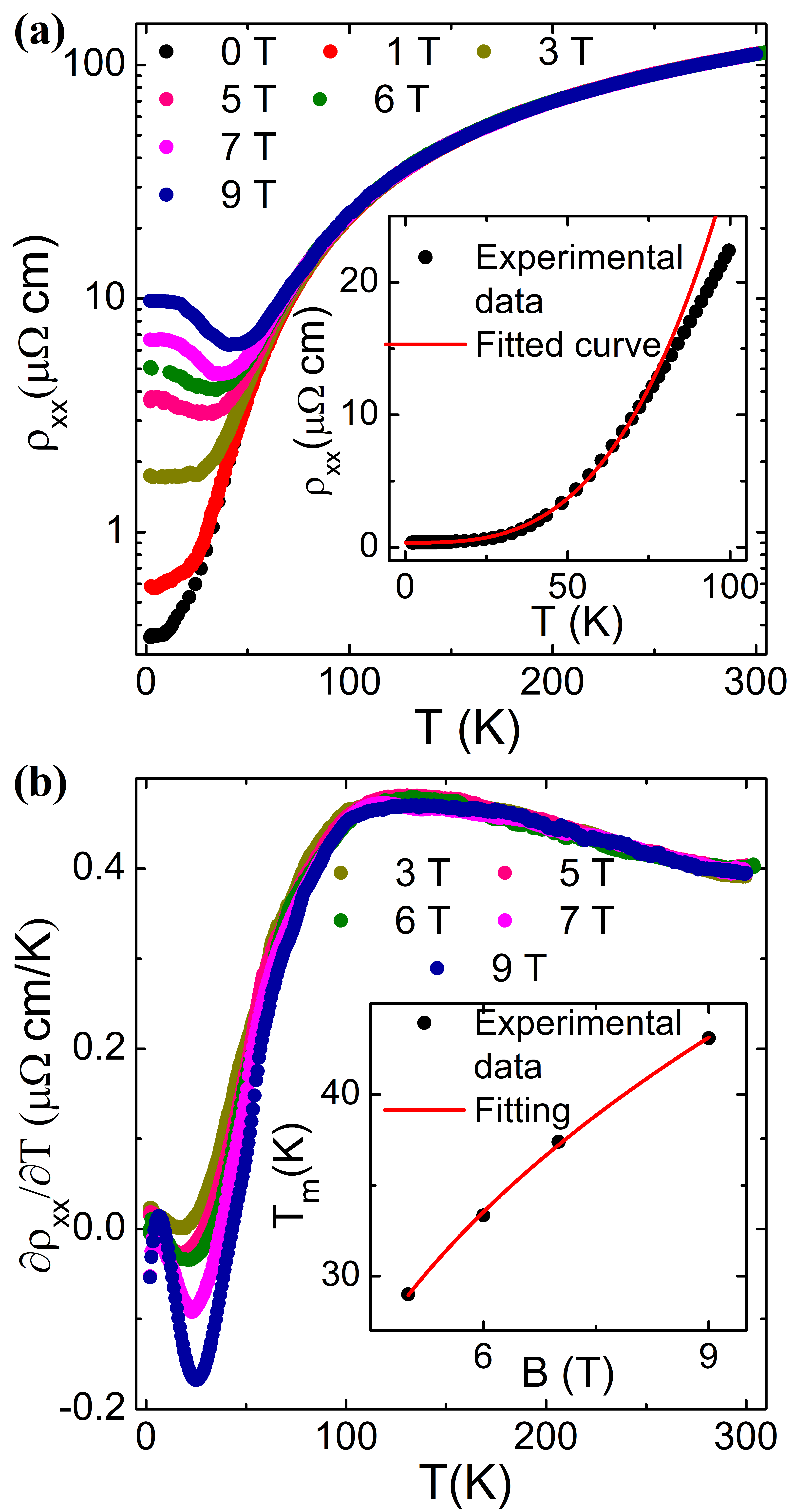}
\caption{(a) Temperature dependence of resistivity at different magnetic field strengths, applied along c-axis. Inset shows the low-temperature region of the zero-field resistivity, fitted using $\rho_{xx}(T)=a+bT^n$ type relation. (b) $\partial\rho_{xx}/\partial T$ vs. temperature curves at different magnetic fields, which reveal two characteristic temperatures $T_{m}$ and $T_{i}$. The field dependence of metal to semiconductor-like crossover temperature ($T_{m}$) is shown in the inset.}
\end{figure}

Next, we have measured the transverse MR of the sample with current along b-axis and magnetic field along c-axis in the temperature range 300 K down to 2 K. At 9 T and 2 K, a very large and non-saturation MR $\sim$2.75$\times$10$^{3}$\% has been obtained [Fig. 3(a)]. The observed value is smaller than other members of the $XPn_{2}$ family \cite{Wang,Shen,Wang2,Li}, but is comparable to several other TSMs \cite{Xiang,Wakeham,Novak,Pavlokiuk}. With increasing temperature the MR decreases drastically and becomes less than 1\% at 300 K. As shown in the inset for a representative temperature, the MR curve obeys almost quadratic magnetic field dependence (MR$\propto$$B^{1.8}$). Keeping the current direction unaltered, when the field is applied along b$\times$c-axis, the MR changes significantly and becomes $\sim$1.1$\times$10$^{3}$\% at 2 K and 9 T. To map the complete directional dependence of the magnetotransport properties, we have rotated the magnetic field in the plane perpendicular to the current, while keeping the current always along b-axis. The calculated MR is shown as a polar plot in Fig. 3(b). The MR shows a two-fold rotational symmetry with 'butterfly-like' strong anisotropy. At 2 K and 9 T magnetic field, the maximum value is $\sim$2.8$\times$10$^{3}$\% at $\sim$170$^{\circ}$ (and $\sim$350$^{\circ}$), whereas the MR becomes as small as 250\% at around 65$^{\circ}$ (and $\sim$245$^{\circ}$). The large anisotropic ratio ($\sim$11.2) indicates the highly anisotropic Fermi surface in MoAs$_{2}$. In addition to this two-fold symmetry pattern, dips and kinks in the MR value have been observed at $\sim$145$^{\circ}$ (and $\sim$325$^{\circ}$) and $\sim$90$^{\circ}$ (and $\sim$270$^{\circ}$), respectively. It may be due to some higher order texturing, which vanishes quickly with increasing temperature. Butterfly-like anisotropic pattern in MR have been previously observed in underdoped cuprate superconductor, manganite system as well as in recently discovered TSM ZrSiS \cite{Jovanovic,Zhang2,Ali2}. However, in these systems, higher order textures are seen to be more robust compared to MoAs$_{2}$. The experimental data can be fitted to a great extent (Fig. 8 in Appendix) assuming contributions from both two-fold and four-fold symmetries \cite{Jovanovic}, MR$(\phi)=C+A_{2}sin[2(\phi-\phi_{2})]+A_{4}sin[4(\phi-\phi_{4})]$, where $A_{2}$ ($A_{4}$) and $\phi_{2}$ ($\phi_{4}$) are the amplitude and phase of the two-fold (four-fold) symmetry, respectively. $C$ is an arbitrary constant. The overall pattern in Fig. 3(b) appears to be tilted with respect to the crystallographic directions. This may be the consequence of the complex Fermi surface and the relative contributions of different Fermi pockets in the transport properties \cite{Collaudin}.

\begin{figure}
\includegraphics[width=0.35\textwidth]{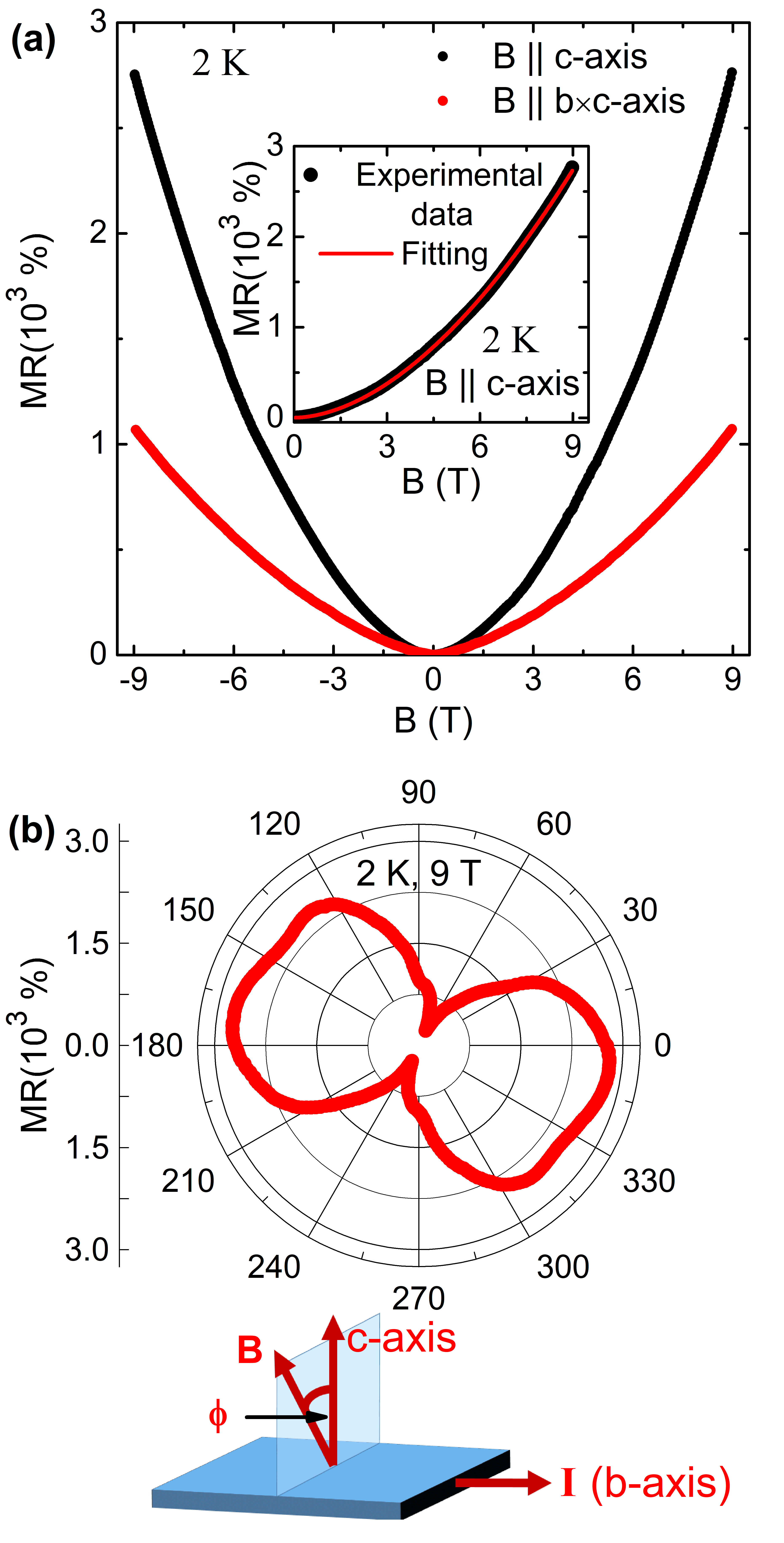}
\caption{(a) Transverse magnetoresistance (MR) of MoAs$_{2}$ at 2 K, when current is along b-axis and magnetic filed is applied along two mutually perpendicular crystallographic directions, b-axis and b$\times$c-axis. The fitting in the inset illustrates nearly parabolic field dependence of MR. (b) Polar plot of MR, when the magnetic field is rotated in the plane perpendicular to the b-axis. The experimental set-up is shown schematically in the inset.}
\end{figure}

We have also measured the MR by changing the angle between current and magnetic field, i.e., by rotating the field along bc-plane. Although the MR becomes minimum with parallel electric and magnetic field (longitudinal configuration) as expected, it still remains positive. In an earlier study \cite{Wang4}, negative LMR has been reported for MoAs$_{2}$ below 40 K. Negative LMR is often considered as a signature of Adler-Bell-Jackiw chiral anomaly \cite{Huang,Li2}, which originates due to the charge pumping between two Weyl nodes of opposite chirality. However, negative LMR can also appear from current jetting effect i.e. due to inhomogeneous current distribution inside the sample \cite{Hu}. In fact, it has been shown for TaAs$_{2}$ that this effect can be easily suppressed by changing the current and voltage lead positions \cite{Yuan}. Nevertheless, chiral anomaly induced negative LMR has a specific temperature and magnetic field dependence \cite{Huang}, which is distinct from current jetting effect. For MoAs$_{2}$, we did not find any negative MR, even when we change the angle between current and magnetic field by few degrees to compensate any possible misalignment.

\begin{figure}
\includegraphics[width=0.5\textwidth]{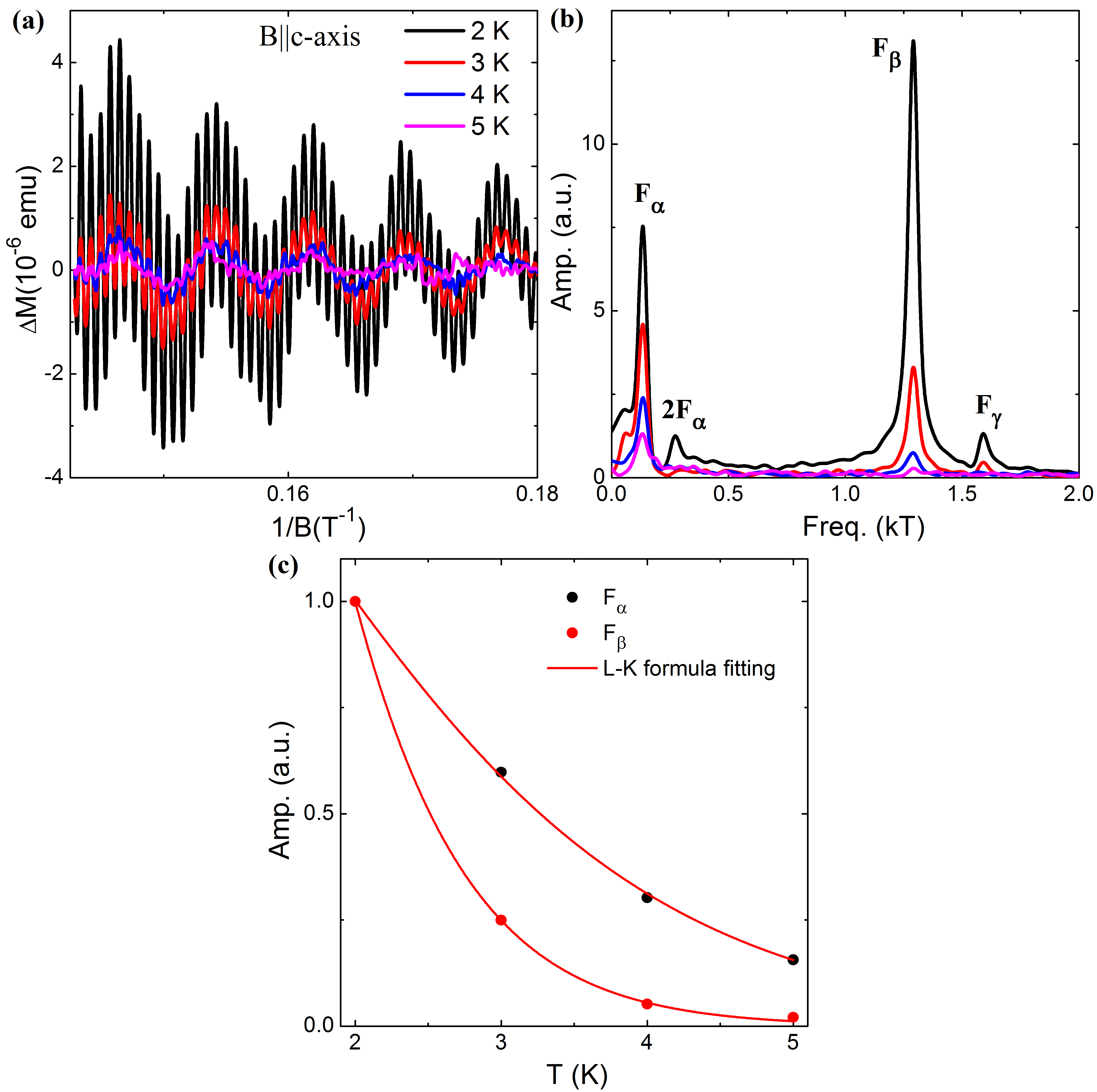}
\caption{(a) de Haas-van Alphen (dHvA) oscillation in the magnetization measurement of MoAs$_{2}$ for magnetic field along c-axis at different representative temperatures. (b) The fast Fourier transform spectra of the dHvA oscillation. (c) Temperature dependence of the oscillation amplitude for two Fermi pockets, $F_{\alpha}$ and $F_{\beta}$.}
\end{figure}

The magnetization measurement of the MoAs$_{2}$ crystals with field along c-axis shows a diamagnetic character with prominent dHvA oscillation up to $\sim$5 K. The oscillatory part in the magnetization data has been extracted by subtracting a smooth background and plotted in Fig. 4(a) as a function of inverse magnetic field at different temperatures. The corresponding fast-Fourier transform (FFT) spectrum in Fig. 4(b), reveals three fundamental frequencies $F_{\alpha}$ [134(3) T], $F_{\beta}$ [1290(2) T] and $F_{\gamma}$ [1589(5) T]. The quantum oscillation frequency ($F$) is related to the extremum Fermi surface cross-section ($A_{F}$) perpendicular to the applied field direction as per the Onsager equation, $F=(\phi_{0}/2\pi^{2})A_{F}$. Here, $\phi_{0}$ is the magnetic flux quantum. Using the relation, we have calculated the Fermi surface cross-sections for these three Fermi pockets perpendicular to the c-axis [See TABLE I]. The calculated cross-sections for $F_{\beta}$ and $F_{\gamma}$, are the largest among the members of $XPn_{2}$ family \cite{Wang,Shen,Yuan,Wang2,Li} as well as other TSMs and comparable to newly discovered three-fold degenerate TSMs WC and MoP \cite{He,Sekhar2}. As a result, the carrier density is also expected to be quite higher than conventional semimetals, which we will discuss latter. The oscillation amplitude decreases rapidly with increasing temperature due to thermal fluctuations and the $F_{\gamma}$ component vanishes completely above 3 K. Therefore, due to lack of experimental points, it is not possible to study the thermal damping of $F_{\gamma}$ component. However, the temperature dependence of oscillation amplitude for $F_{\alpha}$ and $F_{\beta}$ pockets, has been shown in Fig. 4(c) and fitted using the thermal damping factor of the Lifshitz-Kosevich (L-K) formula,
\begin{equation}
R_{T}=\frac{(2\pi^{2}k_{B}T/\beta)}{sinh(2\pi^{2}k_{B}T/\beta)},
\end{equation}
where $\beta=e\hbar B/m^{\ast}$. From the fitting parameters, the cyclotron effective mass ($m^{\ast}$) has been calculated to be 0.37(1)$m_{\circ}$ and 0.74(2)$m_{\circ}$ for $F_{\alpha}$ and $F_{\beta}$ Fermi pockets, respectively. We have also calculated Fermi momentum ($k_{F}$) and Fermi velocity ($v_{F}$) from the quantum oscillations. All the calculated parameters have been summarized in TABLE I.

\begin{center}
\begin{table}
\caption{Fermi surface parameters extracted from dHvA oscillations.}
 \begin{tabular}{|c c c c c c c|}
 \hline
  & Configuration & F & $A_{F}$ & $k_{F}$ & $m^{\ast}$ & $v_{F}$\\

  & & T & 10$^{-3} {\AA}^{-2}$ & 10$^{-3} {\AA}^{-1}$ & $m_{0}$ & 10$^{5}$ m/s\\ [0.5ex]
 \hline\hline
  & B$\parallel$c-axis & 134(3) & 12.8(3) & 63.8(8) & 0.37(1) & 1.99(8)\\
  & & 1290(2) & 123.0(2) & 197.8(2) & 0.74(2) & 3.09(8)\\
  & & 1589(5) & 151.5(4) & 219.6(3) & - & - \\
 \hline\hline
  & B$\parallel$b-axis & 1330(3) & 126.8(3) & 200.9(2) & - & - \\
  & & 1524(5) & 145.3(5) & 215.1(3) & - & - \\
 \hline\hline
  & B$\parallel$b$\times$c-axis & 135(3) & 12.9(3) & 64.0(7) & - & -\\
  & & 458(4) & 43.7(4) & 117.9(5) & - & - \\
  & & 1909(3) & 182.0(3) & 240.7(2) & - & - \\
 \hline
\end{tabular}
\end{table}
\end{center}

dHvA oscillation has also been observed, when the magnetic field is applied along other two mutually perpendicular directions, i.e., along b-axis and b$\times$c-axis (Appendix Fig. 9). From the comparison of the FFT spectrum along different directions, it is clear that the Fermi surface of MoAs$_{2}$ is quite complex with at least three Fermi pockets existing in the system. The Fermi surface geometry also has a large anisotropy, which is consistent with the observed anisotropic MR. The extracted frequency components and related parameters are listed in TABLE I. Note that we could not calculate the effective mass and Fermi velocity of the carriers for the Fermi pockets along these two directions due to lack of experimental data, as the oscillations suppress rapidly with increasing temperature.

\begin{figure}
\includegraphics[width=0.35\textwidth]{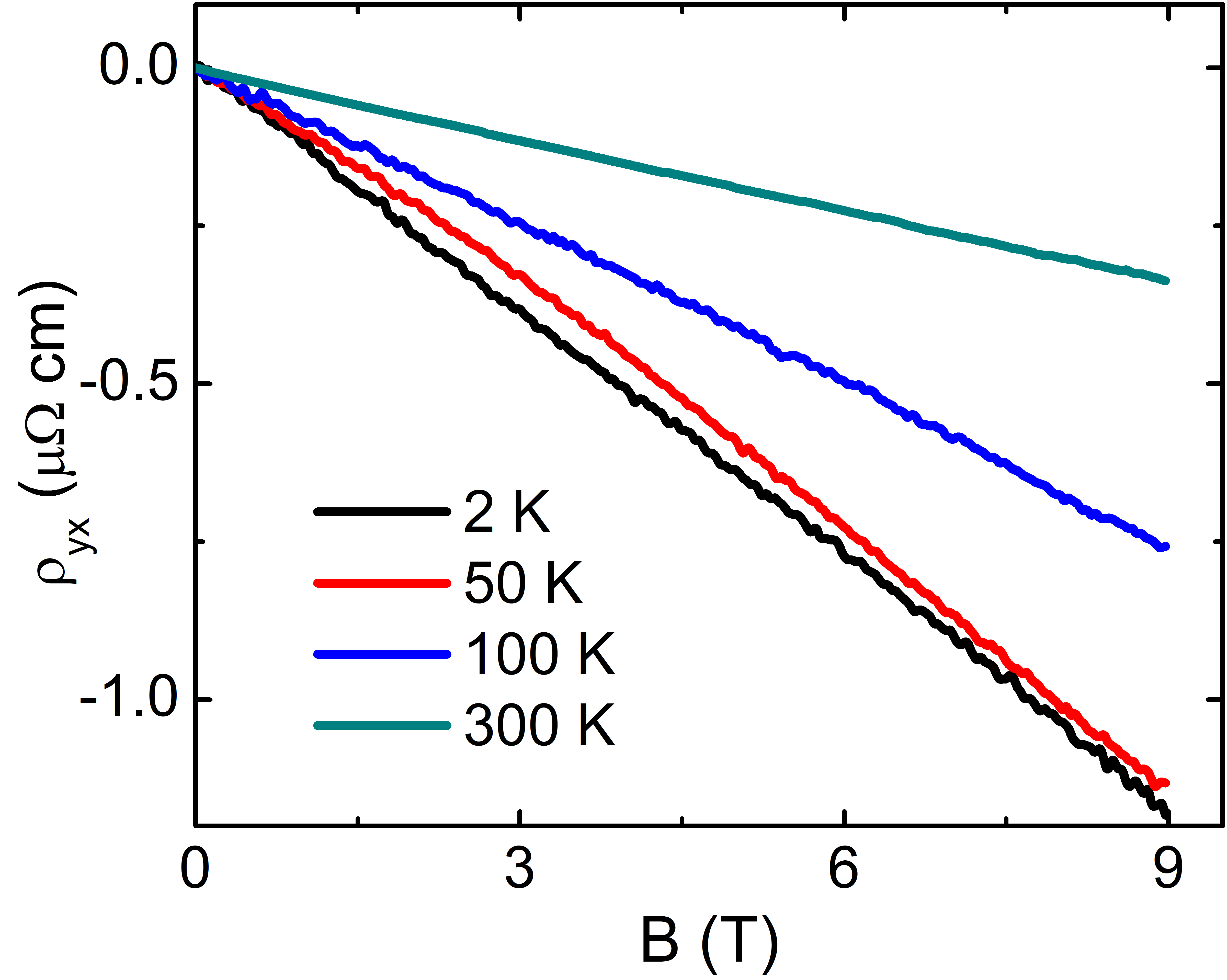}
\caption{Field dependence of Hall resistivity ($\rho_{yx}$) at different temperatures.}
\end{figure}

To get an estimate about the carrier density and mobility, we have measured the Hall resistivity [$\rho_{yx}$] of the crystal at different temperatures. In Fig. 5, $\rho_{yx}$ is shown at some representative temperatures. Hall resistivity is linear and negative at 300 K and indicates electron type charge carriers. As temperature is decreased, $\rho_{yx}$ shows a slightly non-linear character at around 50 K, which is a signature of multi-band transport, as also apparent from our quantum oscillation analysis. However, $\rho_{yx}$ remains negative throughout the measured temperature range and can be described well using one band model. From the slope of the curve $R_{H}=-\frac{1}{ne}$, the electron density ($n$) is calculated to be 4.8(2)$\times$10$^{21}$ cm$^{-3}$ at 2 K. The obtained density is at least two orders of magnitude higher than most TSMs and is in accordance with the large Fermi pockets, which have been observed from the quantum oscillation measurements. In spite of such high carrier density, the carrier mobility is quite large 3.6(1)$\times$10$^{3}$ cm$^{2}$V$^{-1}$s$^{-1}$ and comparable to different topological systems \cite{Novak,Pavlokiuk,Luo}.

The calculated electronic structure of MoAs$_{2}$ is plotted in Fig. 6 along the high-symmetry directions, without considering the effect of spin-orbit coupling. Even with SOC, the band structure remains almost the same. With a larger value of U and with SOC, we find some signature of a magnetic phase. The electronic structure in the low-energy spectrum is dominated by Mo-\textit{d} electrons. The overall electronic structure gives two large Fermi surface pockets, one electron pocket, extending Y and F-points, and a hole-Fermi-surface along the direction of $\Gamma$ to X. With small tuning of the chemical potential, one can find a very tiny hole pocket along $\Gamma$ to Y, and another electron pocket at X-points. However, these two tiny Fermi pockets are subjected to the details of the electronic structure calculation, and more importantly, they are difficult to detect. From our quantum oscillation results, we have identified at least two large and one small Fermi pockets in MoAs$_{2}$. On the other hand, from the Hall measurement, it is clear that the measured crystals are electron doped. Therefore, it is possible that there are one large and another tiny electron pocket, whereas the other large Fermi pocket is hole-type. However, the mobility of the holes must be quite small compared to the electrons so that the transport is dominated only by electron-type charge carriers.

\begin{figure}
\includegraphics[width=0.45\textwidth]{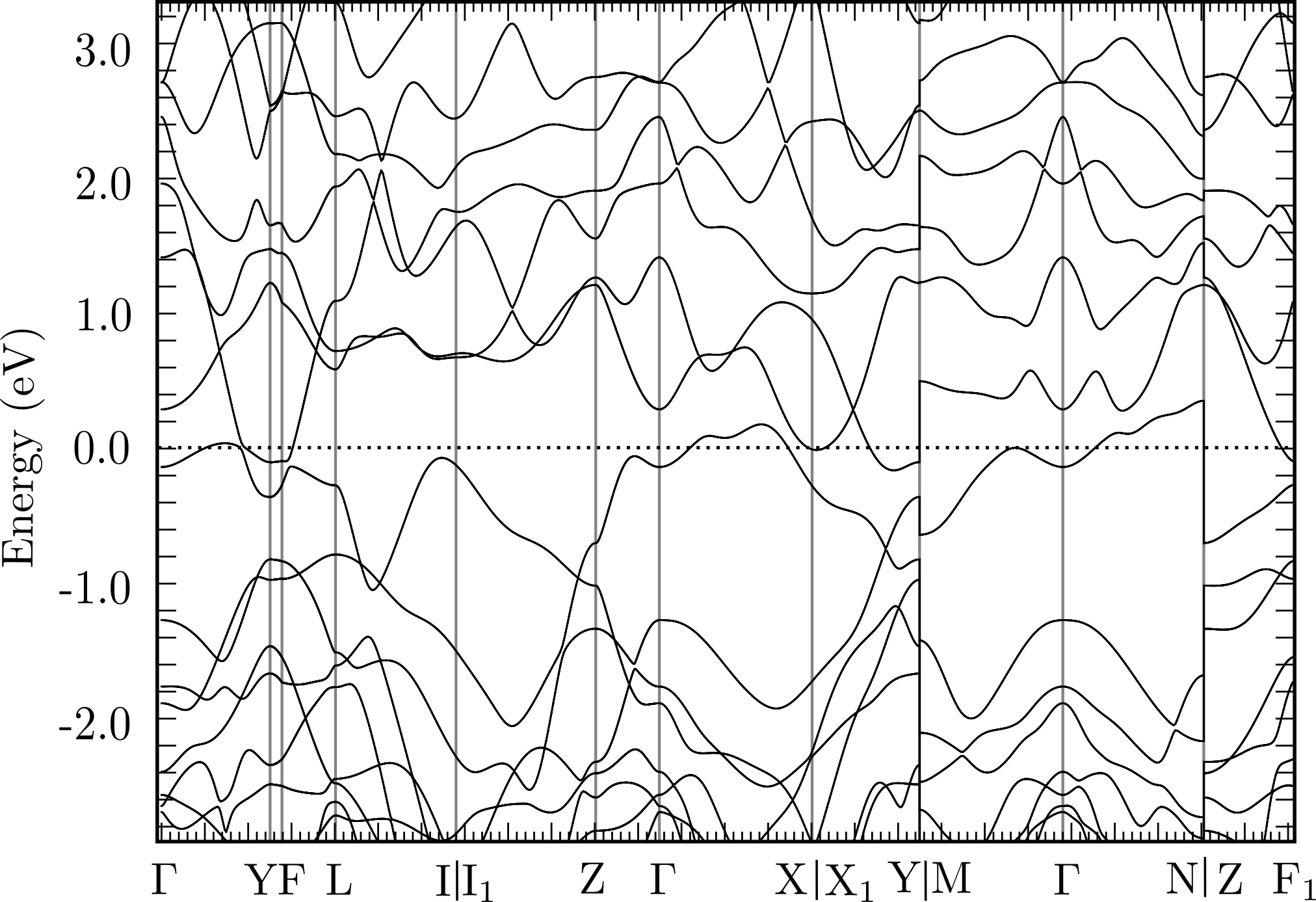}
\caption{The bulk band structure of MoAs$_{2}$ without spin-orbit coupling. The Fermi energy has been aligned with E=0 eV.}
\end{figure}

One of the stringing aspects of this material is that the hole pocket is an 'open-orbit' Fermi surface and is strongly three-dimensional. This leads to an important question: can an open orbit Fermi surface give quantum oscillations? We note that open Fermi surface is also present in several organic compounds \cite{Uji,Senzier,Ohmichi,Weiss,Chaikin,Kornilov1,Kornilov2}, as well as in the Ortho-II phase of YBCO cuprates \cite{Leyraud,Singleton,Sebastian1,Sebastian2,Yelland}. In both these materials, quantum oscillation is observed with multiple frequencies. There are also several theories, modeling the origin of quantum oscillation from open Fermi surface \cite{Varma,Start,Pippard1,Pippard2,Azbel,Osada}. Such open-orbit Fermi surface can also be a possible origin of such large MR in MoAs$_{2}$, as has been earlier proposed for layered compound PdCoO$_{2}$ \cite{Takatsu}.

\begin{figure}
\includegraphics[width=0.45\textwidth]{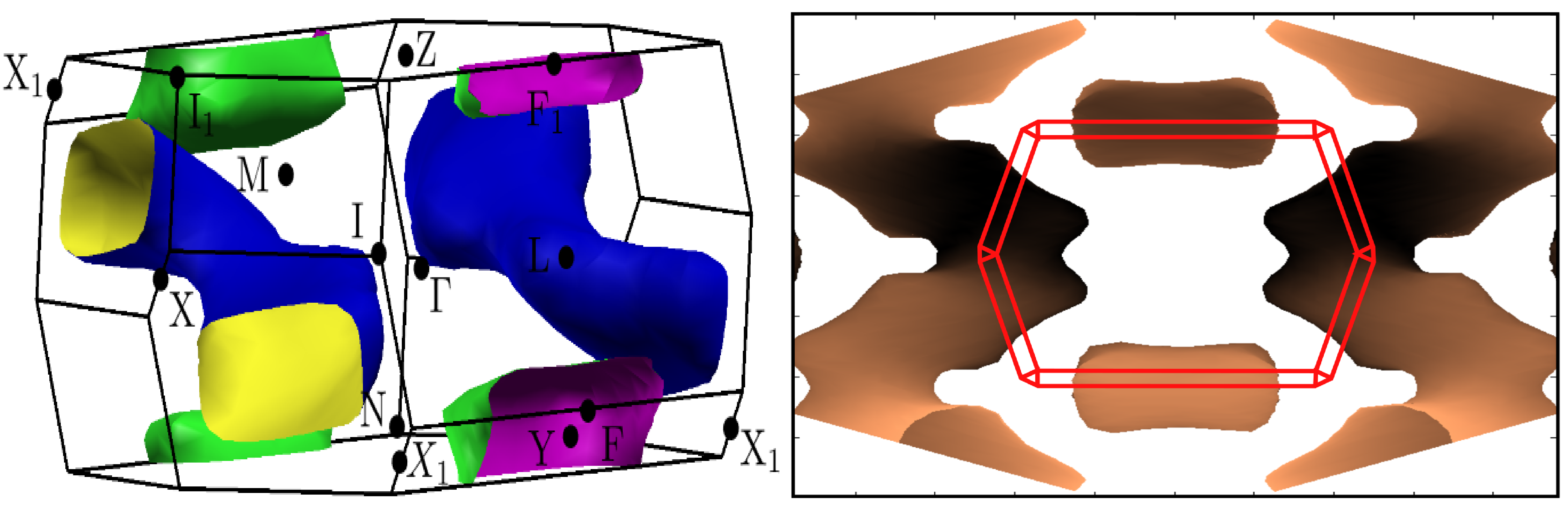}
\caption{The calculated Fermi surface for MoAs$_{2}$, showing two large Fermi pockets including hole-type open-orbit Fermi surface.}
\end{figure}

Unlike in NbAs$_{2}$ and TaAs$_{2}$ materials, MoAs$_{2}$'s electronic structure does not show any Weyl or Dirac cones, at least not a cone which is isolated from other quadratic bands. Even for NbP, a widely studied Weyl semimetal, magnetoresistance measurements demonstrated the coexistence of linear dispersion of a Weyl cone, and quadratic dispersions. The interplay between the two dispersions led to a complex experimental situation, in that while some experiments claim the existence of non-trivial Berry curvature \cite{Shekhar,Wang5}, others found a trivial topological phase in NbP \cite{Sudesh}. In MoAs$_{2}$, there might be a signature of linear band crossing along X1 to Y direction, with the suspected Weyl cone residing about 200 meV above the Fermi level.

\section{Conclusions}

To summarized the results, in this report, we have presented the systematic study of the magnetotransport and Fermi surface properties of a transition metal dipnictide (TMDs) MoAs$_{2}$. Field-induced resistivity plateau and a large magnetoresistance have been observed, which are two of the characteristic features of a topological semimetal. The magnetoresistance shows a strong butterfly-like anisotropy, when the magnetic field is applied along different crystallographic directions. Unlike other isostructural compounds, Hall measurements for MoAs$_{2}$ reveal quite high carrier density with electron-type majority carriers. The Fermi surface has been probed by de Haas-van Alphen oscillation along three mutually perpendicular directions. The observed Fermi pockets are largest among the members of the $XPn_{2}$ family and are highly anisotropic in nature. The first-principles calculations show that the electronic band structure of MoAs$_{2}$ is significantly different from that for other TMDs. The Fermi surface mainly consists of one large electron pocket and an open-orbit hole pocket, which may be the origin of such high magnetoresistance in this compound.

\section{Acknowledgements}
We like to thank A. Pal and S. Roy for their help during experimental measurements. TD acknowledges the financial support from the Department of Science and Technology (DST), India under the Start Up Research Grant (Young Scientist) [SERB No: YSS/2015/001286].

\section{Appendix}
\newpage
\begin{figure}
\includegraphics[width=0.25\textwidth]{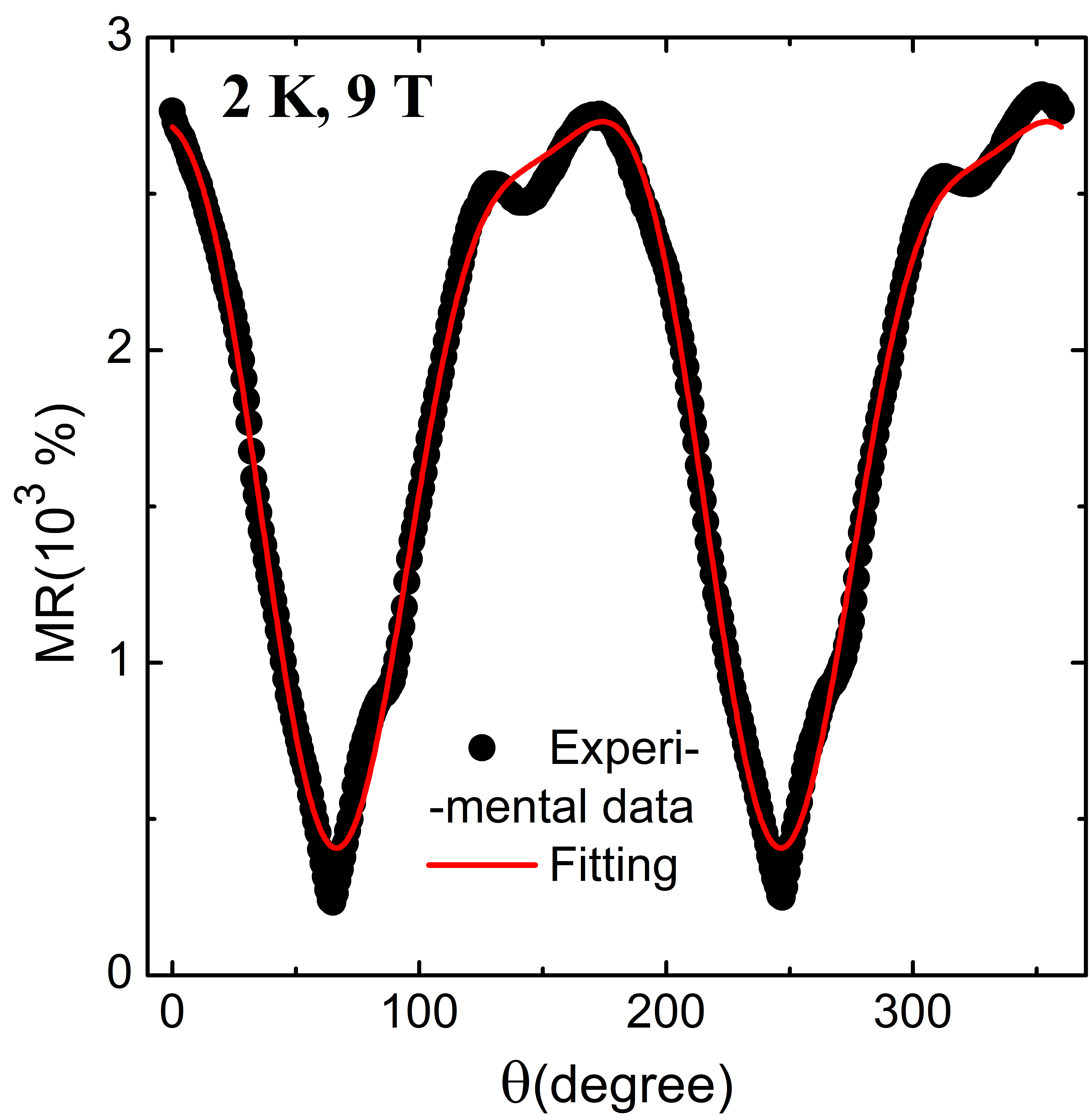}
\caption{Angle dependence of the transverse magnetoresistance, fitted assuming both two-fold and four-fold symmetry contributions.}
\end{figure}

\begin{figure}
\includegraphics[width=0.3\textwidth]{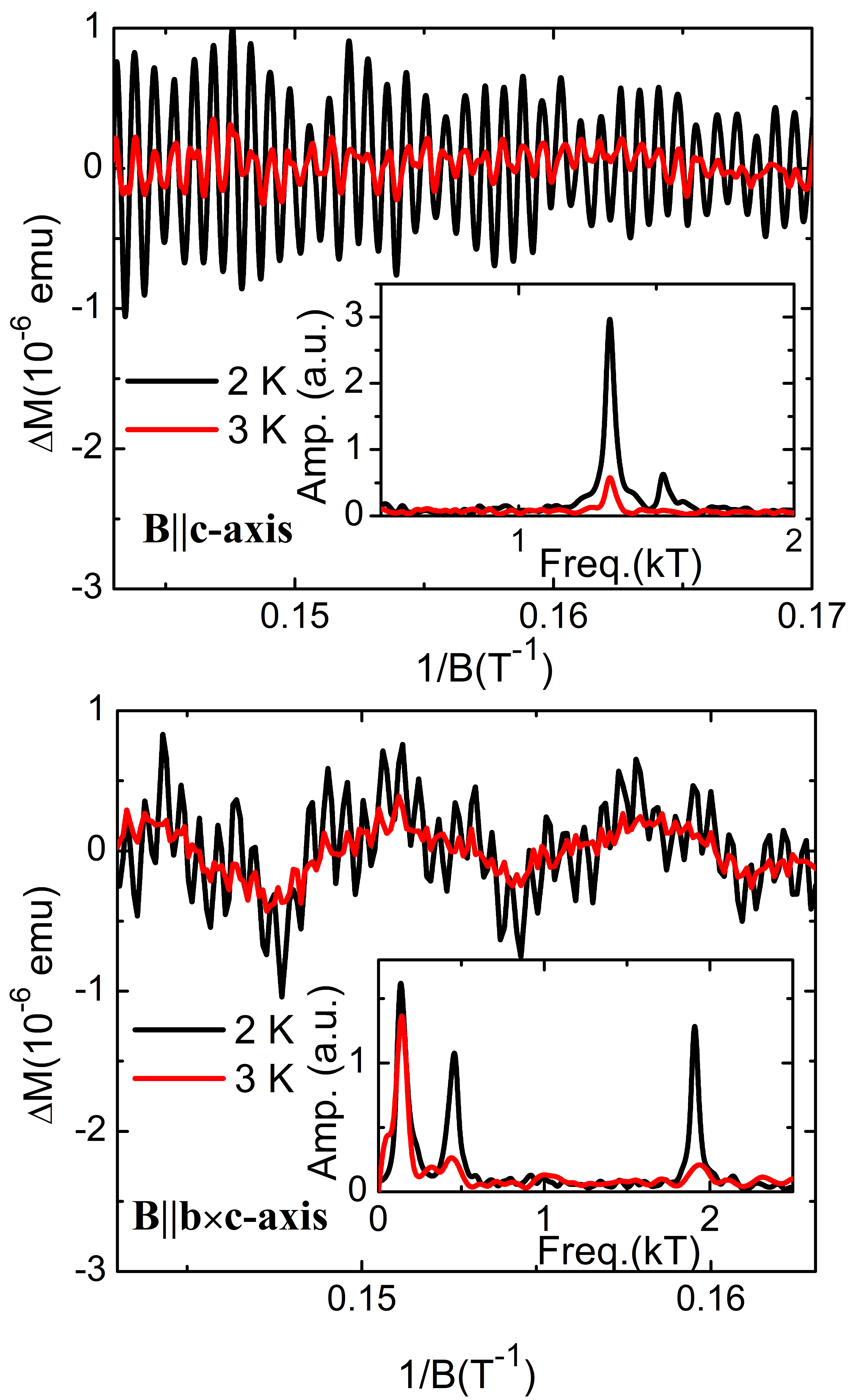}
\caption{de Haas-van Alphen (dHvA) oscillation for magnetic field along b- and b$\times$c-axes. Insets show the corresponding FFT spectrum.}
\end{figure}

\end{document}